\documentclass{PoS}

\usepackage{nccmath}
\usepackage{array,ragged2e}

\title{The effect of an infrared divergent quark-antiquark interaction kernel 
on other Green functions}

\ShortTitle{The effect of an IR divergent $q\bar q$ interaction 
kernel on other Green functions}

\author{\speaker{Reinhard Alkofer} \\
Institut f\"{u}r Physik, Karl-Franzens-Universit\"{a}t Graz, 
Universit\"{a}tsplatz 5, 8010 Graz, Austria\\
E-mail: \email{reinhard.alkofer@uni-graz.at}}

\author{Mario Mitter\\
        Universit\"{a}t Heidelberg, Institut f\"{u}r Theoretische Physik, 
	Philosophenweg 16, D-69120 Heidelberg, Germany\\
        E-mail: \email{mario.mitter@thphys.uni-heidelberg.de}}

\abstract{
The $n$-point Green functions of Landau gauge QCD are systematically investigated
in a Dyson-Schwinger approach assuming a static linearly rising potential between
fundamental color charges. Besides quarks also scalar matter in the fundamental
representation is considered. Starting from the hypothesis of an $1/k^4$ infrared
divergent matter-antimatter vertex restrictions on the general color tensor
structure of this divergence are derived. Consequences for the other four-point
functions of QCD, resp., scalar QCD, are presented. Hereby Casimir scaling is 
found. It is shown that possible
singular contributions to the three-point functions vanish due to cancellations
within the color algebra. On the other hand, higher $n$-point functions inherit 
the infrared singularity of the matter-antimatter vertex  in certain color
channels. The presented results show that linear confinement is consistently
possible in a Greens function approach, however, at the expense that the 
decoupling theorem is circumvented by infrared singularities. 
}

\FullConference{Xth Quark Confinement and the Hadron Spectrum\\
		 8-12 October 2012\\
		 TUM Campus Garching, Munich, Germany}

\begin{document}

\section{Motivation}

The title of this conference series is {\it Quark Confinement and the Hadron
Spectrum}. Over the last eighteen years this conference series has documented the
impressive progress on non-perturbative QCD and hadron physics which has been
achieved by a wealth of methods. Nevertheless, it is fair to say that our current
understanding of quark confinement is far from being satisfactory. In this talk
another facet is added to the discussion. It is based on the so-called quenched
approximation and the then resulting linearly rising static potential in between
infinitely heavy, {\it i.e.}, non-dynamical, quarks. Such a potential has been
impressively verified by numerous lattice calculations in which the Wilson loop  has
been computed and its area law related to the linearly  rising potential has been 
extracted with high precision. 

\begin{figure}[b]
\begin{center}
\begin{minipage}{0.4\textwidth}
\includegraphics[width=0.9\textwidth]{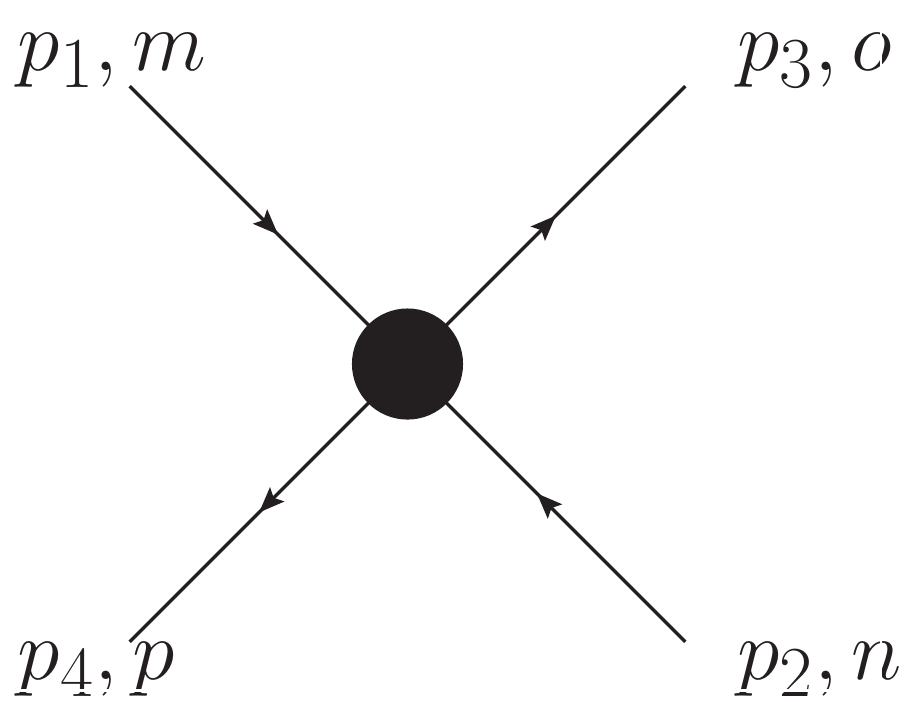}
\end{minipage}
\begin{minipage}{0.3\textwidth}
{ \Large
$\displaystyle{
\left. \stackrel{p_1\to p_3}{\sim} 
\frac {\delta_{mo} \delta_{np}} {(p_1-p_3)^4} \right|_{\mathrm
{reg.}}}$}
\end{minipage}
\end{center}
\caption{\label{fig1} An illustration of the color indices and the momentum
dependence of the matter four-point function as well as the color tensor
structure for the assumed infrared singularity.}
\end{figure}

In this talk the question whether and how such a linearly rising static potential
can be encoded in the $n$-point Greens functions of quenched QCD will be
discussed. Although the corresponding investigation is still far from being
complete some interesting positive answers along with some puzzling facts will 
result. But as a first step a disclaimer is in order.  Due to the exponentiation
of the gluon field the Wilson loop depends on infinitely many $n$-point
functions.  Therefore the observed area law of the Wilson loop does not provide
a  compelling reason why a finite set of $n$-point functions should already lead
to confinement in the sense of a linearly rising potential. On the other hand,
one can show that an infrared singular quark interaction can provide such a
linearly rising potential. The typical starting point for such an investigation
(also  taken here) is the hypothesis that some tensor components of the quark
four-point function diverges like $1/k^4$ for small exchanged momentum $k$ ({\it
cf.}, Fig.~\ref{fig1}). If such an infrared divergence is properly 
regularized~\cite{Gromes:1981cb}
and then Fourier transformed  it leads, in the non-relativistic limit, to a heavy
quark potential with a term linear in the distance $r$,  {\it i.e.}, to the
anticipated linearly rising potential. This provides an example how
confinement can be encoded already in a single $n$-point function. 

In the following we will only briefly summarize the main line of argumentation
and then the consequences for other Greens functions which are implied by an
infrared singular four-point function, a detailed discussion can be found in
Ref.~\cite{Mario}.

\section{A remark on Landau gauge QCD Greens functions}

As Greens functions of colored fields are gauge dependent quantities one might
even speculate whether in different gauges the situation is qualitatively
different. Of course, as we wish to investigate confinement in a functional
approach we would choose a gauge where confinement is contained in a finite set
of $n$-point functions. As by now there is no hint which gauge might be best
suited for our planned study we simply choose the gauge in which the Greens
functions are best known, the Landau gauge. Also in this respect we
cannot be sure that in the non-perturbative domain Landau gauge is uniquely
defined, see, {\it e.g.}, Ref.\ \cite{Maas:2013vd} and references therein.

Actually, early investigations of the Landau gauge gluon propagator were
exploring the possibility whether the $1/k^4$ infrared singularity might already
exist in this propagator, see, {\it e.g.}, Refs.\
\cite{Mandelstam:1979xd,Brown:1988bn} and references therein. By now it is clear
that the Landau gauge gluon propagator is suppressed in the infrared and not
enhanced\footnote{A discussion of the infrared behaviour of the Landau gauge
gluon propagator can be found, {\it e.g.}, in the reviews  \cite{Alkofer:2000wg}
and references therein. In other gauges one might have infrared enhanced and
infrared suppressed components of the gluon propagator. Examples are Coulomb
gauge with an infrared enhanced temporal and an infrared suppressed spatial gluon
propagator, or maximally Abelian gauge with an infrared enhanced color-diagonal
and an infrared suppressed  color-off-diagonal gluon propagator.}: 
The Landau gauge gluon is confined and not confining. On
the other hand, in the so-called scaling solution of Dyson-Schwinger and
Functional Renormalization Group Equations the quark-gluon vertex can be infrared
singular such that the four-point function assumes the $1/k^4$ singularity 
\cite{Alkofer:2008tt}. In this respect it is interesting to note that such an 
infrared singularity provides a description of the $U_A(1)$ anomaly within a
Greens function approach \cite{Alkofer:2008et}. Taken all this together motivates
our study of the possibility of an $1/k^4$ singularity in the matter four-point
function in Landau gauge QCD as well as an investigation of the consequences for
all other Greens functions.\footnote{Although hadrons as colorless bound states
are in the focus of interest when studying the four-point functions in the meson
and the six-point functions in the baryon channels there is also some reason to 
study colored higher $n$-point functions, {\it e.g.}, to look for negative-metric
and zero-norm bound states containing ghosts to complete the non-perturbative
BRST quartets \cite{Alkofer:2013ih}.}

\section{Assuming an infrared divergent four-point function}

As detailed above the hypothesis to be tested in this investigation is whether
the four-point function can be maximally infrared singular without causing any
contradiction when plugged in into the equations for an arbitrary Greens
function. Here we will focus on the use of Dyson-Schwinger equations for this
purpose. As the derivation of Dyson-Schwinger equations for higher $n$-point
functions by hand ranges from tedious to impossible corresponding tools have been
employed \cite{Huber:2011qr}. A good example for the complexity of such a
Dyson-Schwinger equation is already the one for the quark four-point function,
see Fig.~\ref{fig2} for a diagrammatical representation.

\begin{fleqn}

{\centering

\begin{figure}[t!]

\begin{tabular}{ m{2.0cm} m{0.5cm} m{2.0cm} m{0.5cm} m{2.0cm} m{0.5cm} m{2.0cm} m{0.5cm} m{2.0cm} m{0.5cm}}
\centering
    \includegraphics[width=2cm]{CS4}
&
$=$
&
\centering
    \includegraphics[width=2cm]{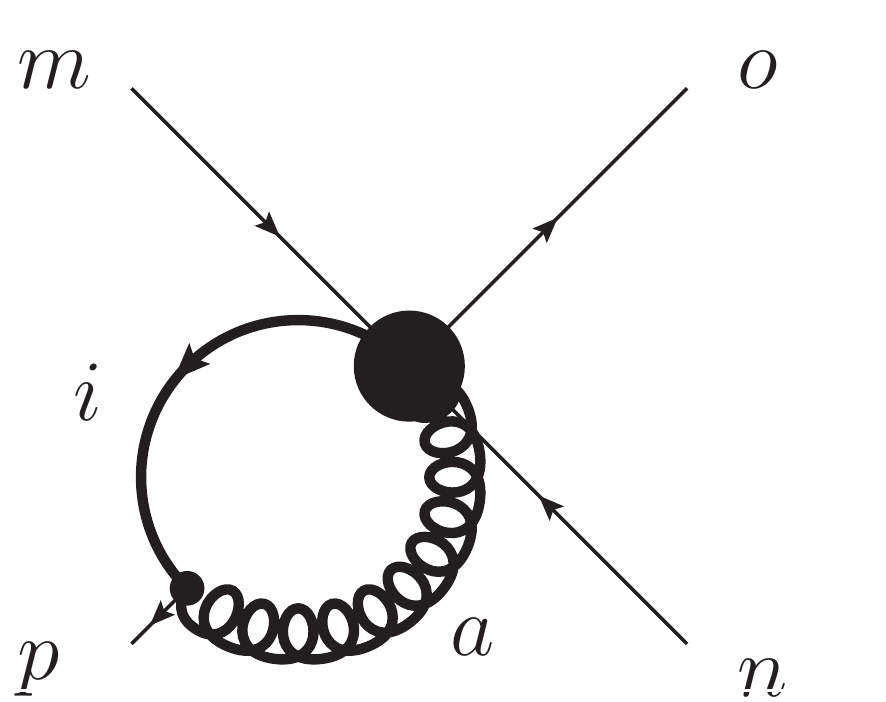}
&
$+$
&
\centering
    \includegraphics[width=2cm]{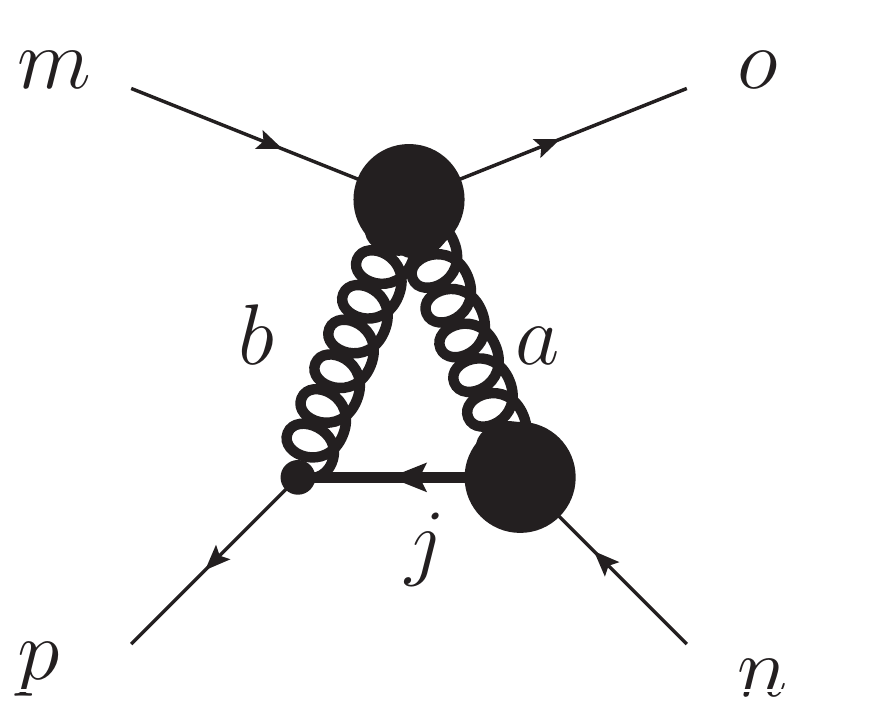}
&
$+$
&
\centering
    \includegraphics[width=2cm]{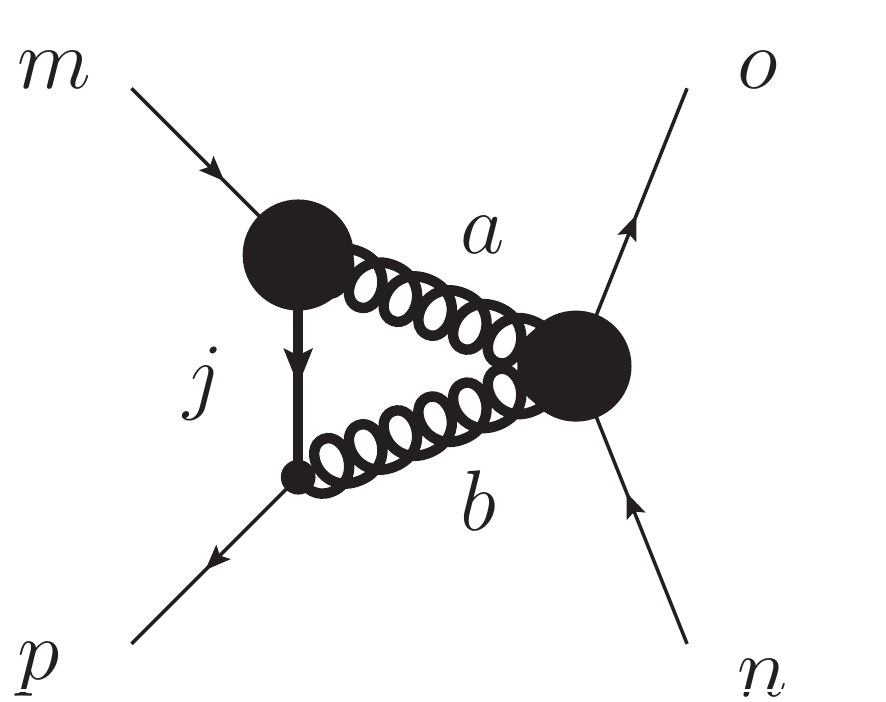}
&
$+$
&
\centering
    \includegraphics[width=2cm]{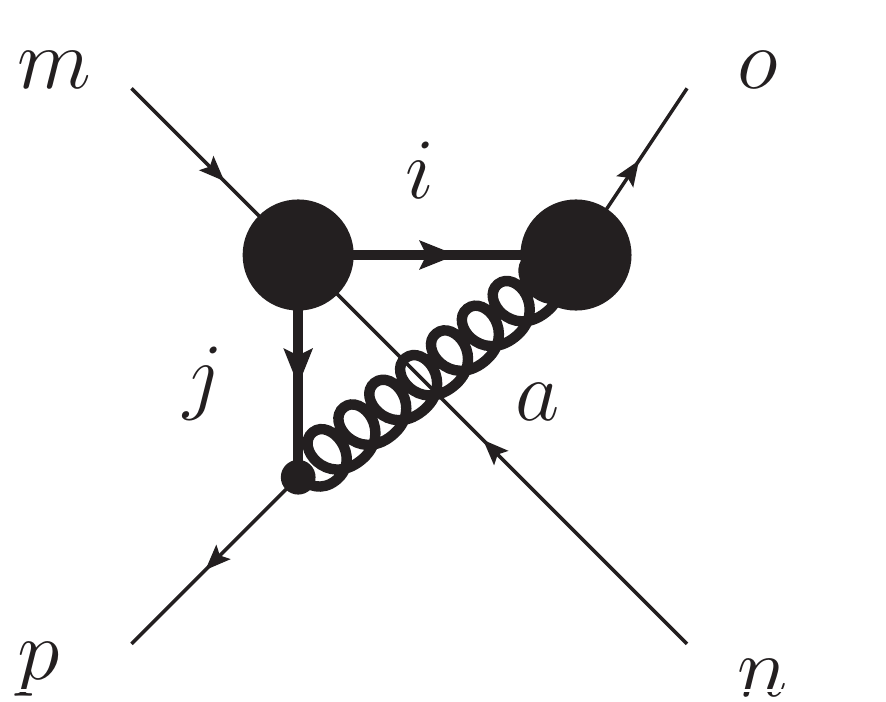}
&

\\

&
$-$
&
\centering
    \includegraphics[width=2cm]{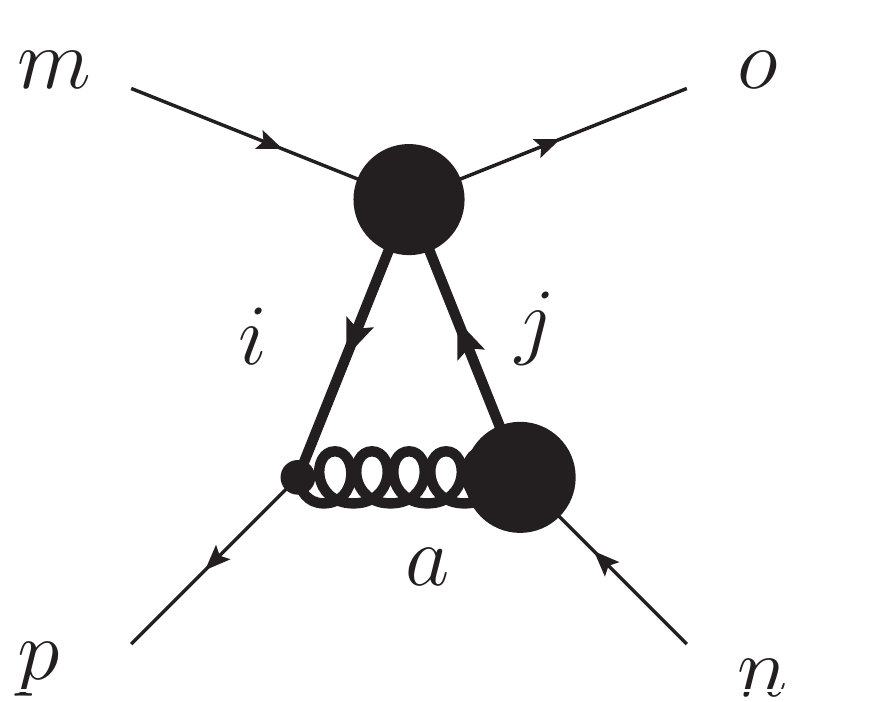}
&
$-$
&
\centering
    \includegraphics[width=2cm]{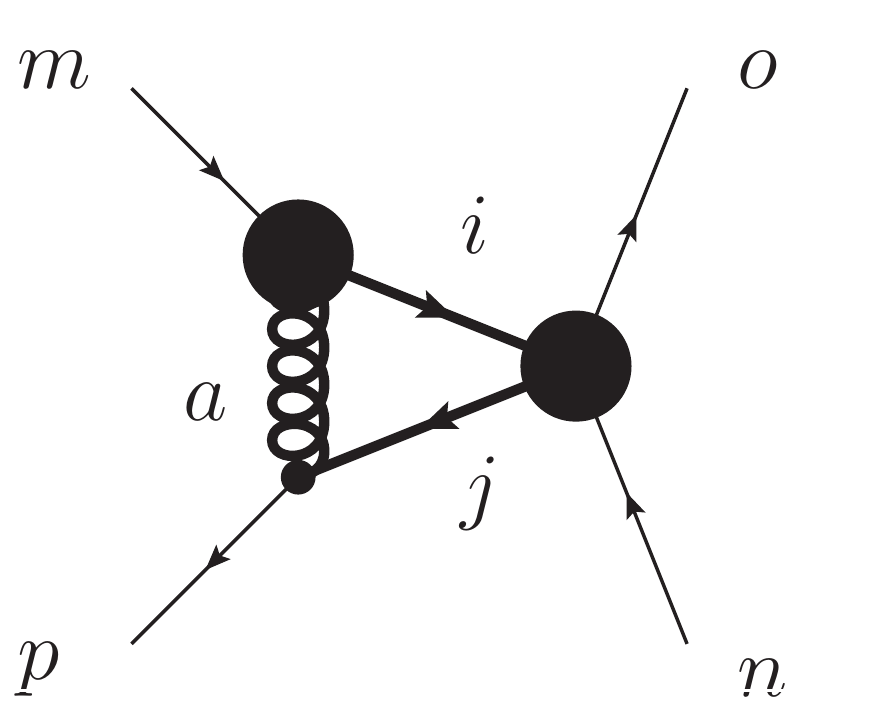}
&
$+$
&
\centering
    \includegraphics[width=2cm]{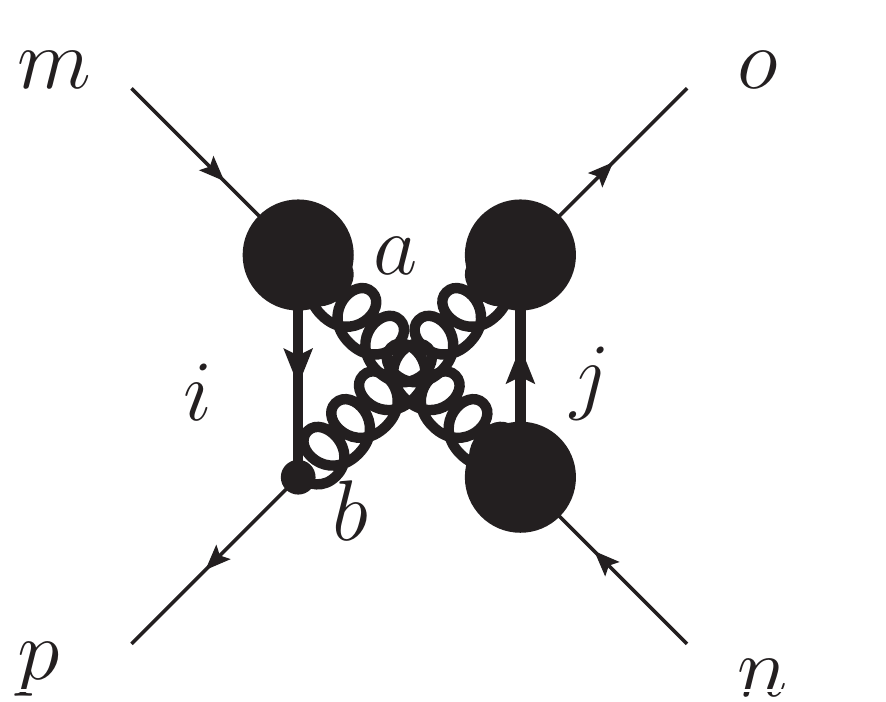}
&
$+$
&
\centering
    \includegraphics[width=2cm]{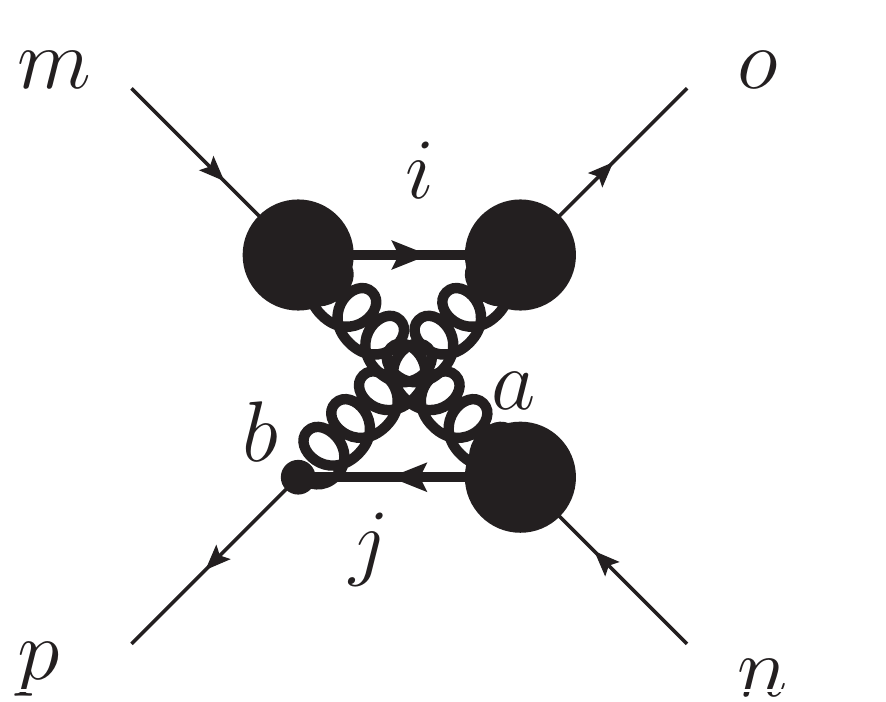}
&

\\

&
$+$
&
\centering
    \includegraphics[width=2cm]{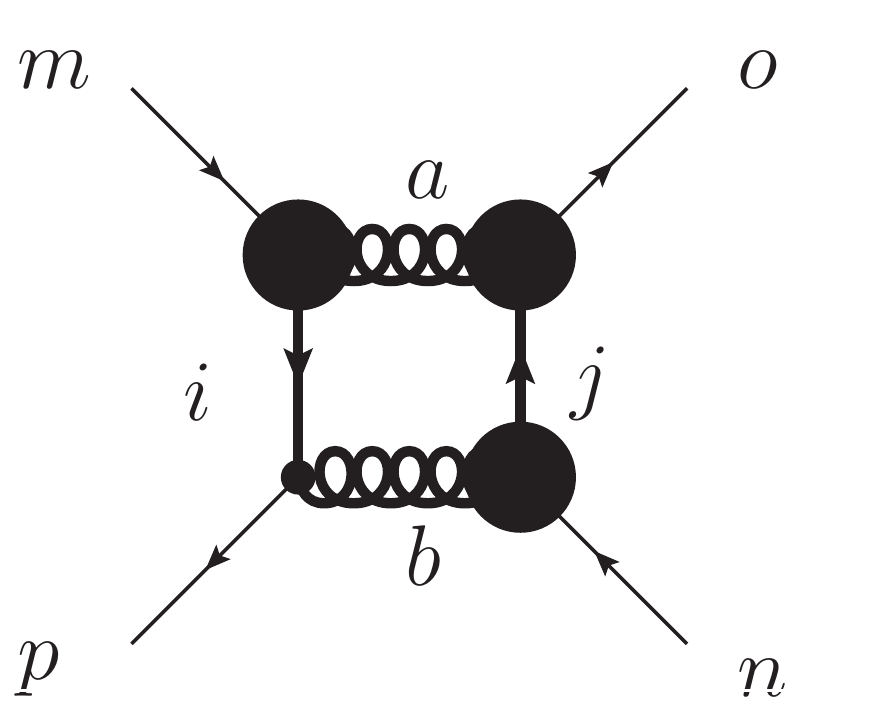}
&
$+$
&
\centering
    \includegraphics[width=2cm]{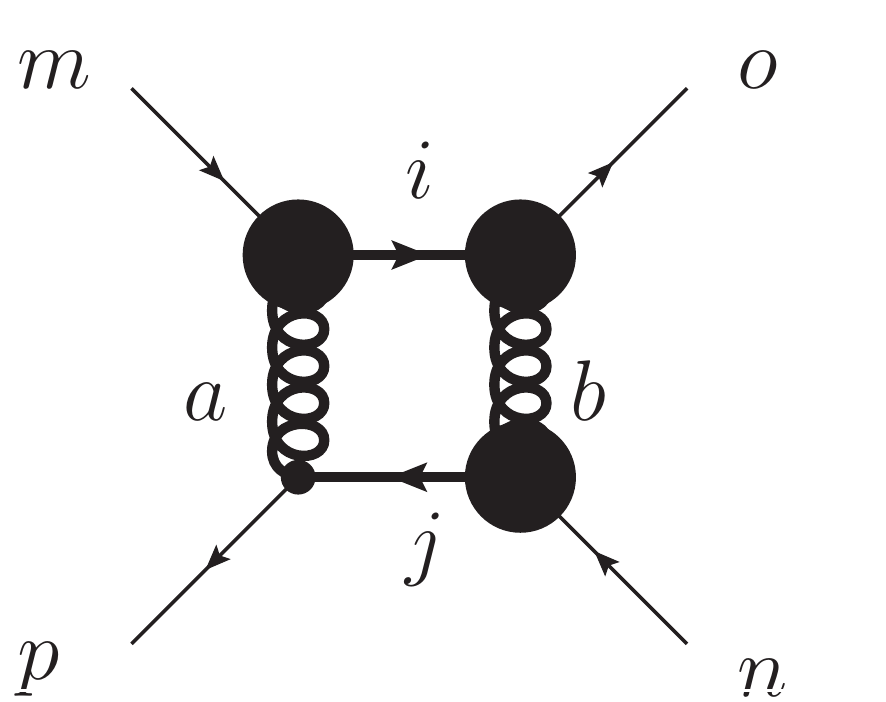}
&

&

&

&

&

\end{tabular}
\caption{The Dyson-Schwinger for the quark four-point function.}
\label{fig2}
\end{figure}

}

\end{fleqn}

As we are interested in the role of infrared divergent terms in the confinement
mechanism special emphasis will be laid on the color structure of potentially
infrared divergent terms. The spinor property of quarks is neglected in the
following discussion as fundamentally charged matter is also expected to
be confined.
In the calculations we employed as gauge group SU$(N)$.

At first sight the Dyson-Schwinger equation for the matter four-point function
causes no immediate contradiction. But a restriction arises when considering the
equation for the matter-gluon vertex which would receive a singular contribution
\begin{fleqn}

{\centering

\begin{tabular}{>{\hfill\centering\arraybackslash} m{7.5cm} >{\arraybackslash}m{7.0cm}}
\centering
    \includegraphics[height=3cm]{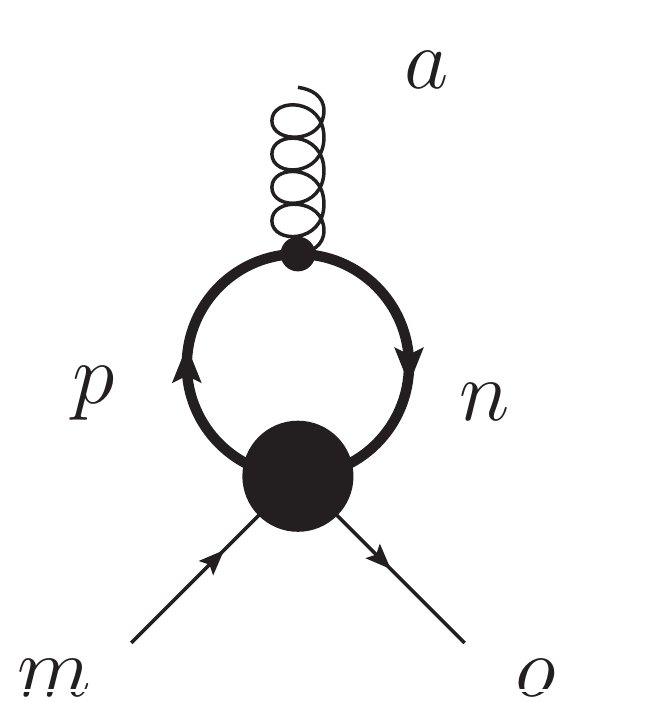}
&
    \begin{eqnarray}\label{eq:quarkglue_vertex_divergency}
     \stackrel{p_m\rightarrow p_o}{\varpropto} \frac{1}{(p_m-p_o)^4}\ .
    \end{eqnarray}
\end{tabular}

}

\end{fleqn}
\noindent Feeding the resulting infrared behavior
of this vertex into the equations for $n\ge 4$ leads to
divergencies stronger than the initially assumed $1/k^4$
behavior. In case of the quark four-point functions the 
infrared singularity of the box diagrams would exceed
the initially assumed divergent behavior by infrared power counting.

Therefore the boundedness of higher $n$-point functions to a maximal infrared singularity
of the $1/k^4$ type requires a less singular
matter-gluon vertex which in turn implies that the infrared singularity can only
be present in one of the two possible color structures, the divergent one
is shown in Fig.~\ref{fig1}. In this respect it is interesting to note that 
{\em one-gluon exchange fails to reproduce this color structure.} The fact that
in this way one can exclude an one-gluon exchange  as source for confinement
comes as no surprise: Such a type of confinement would lead to the analogue of
van-der-Waals forces \cite{Greensite:2011zz} which are not seen in nature.

By identifying corresponding diagrams one can more or less straightforwardly
demonstrate that in case of a $1/k^4$ singularity in the color non-singlet
channel of the matter four-point function all other four-point functions ({\it
i.e.}, the four-gluon vertex as well as the ghost-gluon, matter-ghost and
matter-gluon scattering kernels) acquire an analogous infrared singularity in
respective specific color channels. For example the quark-gluon four-point function
receives a contribution 
\begin{fleqn}

{\centering

\begin{tabular}{>{\hfill\centering\arraybackslash} m{6.5cm} m{0.5cm} >{\arraybackslash}m{7.1cm}}
\centering
    \includegraphics[height=3cm]{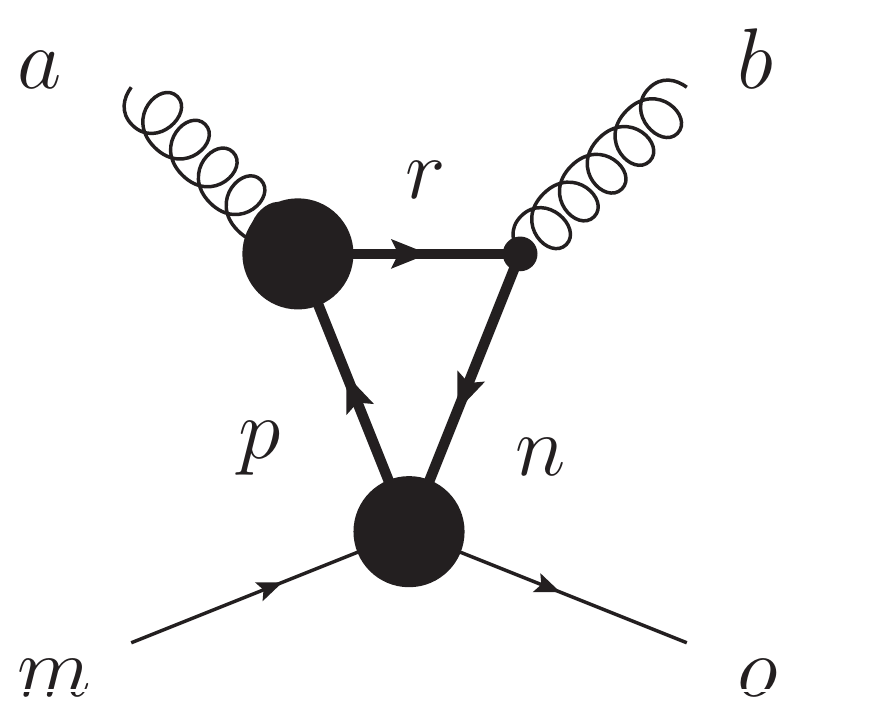}
&
:
&
    \begin{equation}\label{eq:matterglue_4pt_singularity}
     \varpropto \frac{\delta_{mo}\delta^{ab}}{(p_m-p_o)^4}\ .
    \end{equation}
\end{tabular}

}

\end{fleqn}
\noindent Again straightforward application of
identities of the SU$(N)$ color algebra can be used to check that these newly
appearing singularities are not forwarded to the quark-gluon vertex or other
three-point functions
\begin{fleqn}

{\centering

\begin{tabular}{>{\hfill\centering\arraybackslash} m{7.5cm} >{\arraybackslash}m{7.00cm}}
\centering
    \includegraphics[height=3cm]{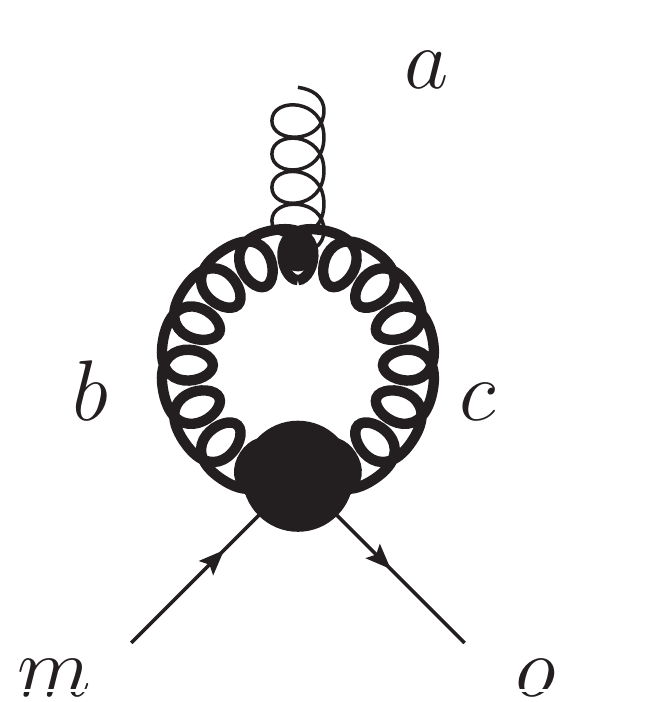}
&
    \begin{equation}
     :\ f^{abc}\delta^{bc}\delta_{mo} =0\ .
    \end{equation}
\end{tabular}

}

\end{fleqn}
Even higher $n$-point functions can be treated when factorizing the tensor
structures with the help of diagrammatic methods \cite{Cvitanovic:2008zz}. This
allows to map the color weights of diagrams of $(n+m)$-point functions to the
color weights of diagrams of $n$- and $m$-point functions. In this way one can
inductively derive how the starting hypothetical $1/k^4$ infrared singularity
in the four-point function propagates through the tower of all higher Greens
functions. Thus one can conclude that if the starting hypothesis is true  all
higher $n$-point functions contain contributions $\propto 1/k^4$ with $k$ being
the momentum transfer between two colored clusters. 

\section{Discussion}

One non-trivial result of this analysis is the fact that propagators and
three-point functions are protected by systematic cancelations in the color
algebra against infrared singularities. These, if existed, would in turn lead to
unaccceptably strong infrared singularities in the $n\ge4$-point functions. We
have verified these cancelations if the color group is SU$(N)$. It would be 
interesting to check whether the algebras of other gauge groups as, {\it e.g.},
SO$(N)$ or $G_2$, allow for the same conclusion.

The second, although not so astonishing,  result which is nevertheless
worth mentioning
 is that an one-gluon exchange
cannot be the (only) source of an infared singularity. As already metioned the
old picture of infrared slavery by gluon exchange is anyhow challenged by
contradictions to phenomelogy and lattice results \cite{Greensite:2011zz}.  

Third, one can check by explicit inspection of the color weights that the assumed
confining $1/k^4$ singularity in the four-point function implies Casimir scaling
(which has been unambigously identified in lattice calculations) for the
confining potential between colored clusters. If for example the singularity
originates from the diagrams containing the quark-gluon four-point function
a simple calculation yields singular contributions
\begin{fleqn}

{\centering

\begin{tabular}{>{\hfill\centering\arraybackslash} m{5cm} m{0.5cm} >{\arraybackslash}m{8.6cm}}
\centering
    \includegraphics[height=3cm]{CA2s2conts4_1}
&
:
&
\begin{eqnarray}\label{eq:A2s2connts4_1}
 T^{a}_{nj}T^b_{jp}\delta^{ab}\delta_{mo} = \frac{N^2-1}{2N}\frac{\delta_{mo}\delta_{np}}{(p_m-p_o)^4}\ ,
\end{eqnarray}
\end{tabular}

}

\end{fleqn}
\begin{fleqn}
{\centering

\begin{tabular}{>{\hfill\centering\arraybackslash} m{5cm} m{0.5cm} >{\arraybackslash}m{8.6cm}}
\centering
    \includegraphics[height=3cm]{CA2s2conts4_2}
&
:
&
\begin{eqnarray}\label{eq:A2s2connts4_2}
 T^{a}_{mj}T^b_{jp}\delta^{ab}\delta_{no} = \frac{N^2-1}{2N}\frac{\delta_{mp}\delta_{no}}{(p_m-p_p)^4}\ .
\end{eqnarray}
\end{tabular}

}

\end{fleqn}
\noindent For a general representation of the gauge group these diagrams result in a singular
contribution proportional to the quadratic Casimir invariant. Similarly other
diagram contributing to the quark four-point function can be either excluded as
sources of the singularity or yield the correct Casimir scaling factor.

 In addition, the color channels of
the infrared singularities in higher $n$-point functions are in accordance with
arguments how the cluster decomposition property might fail in QCD. Proper
regularization of the $1/k^4$ singularity in the  higher $n$-point functions
results in a  $\delta$-function type  contribution which in turn relates to a
possible way of violating  cluster decomposition.

Fourth, and actually most surprising, the presented results demonstrate how the
decoupling theorem \cite{Appelquist:1974tg}
is circumvented by infrared singularities: 
One very heavy fundamental charge, once introduced into the theory, will by
virtue of the $1/k^4$ infrared singularity in its four-point function induce a 
change in the infrared behavior of Greens functions in the Yang-Mills sector. 

\section{Summary and Outlook}

To summarize, we remark that the assumption of a confining infrared singularity
in the matter-matter scattering kernel leads to several wanted features.
Certainly not all possible consequences of a $1/k^4$ singularity in the quark
four-point function  have been investigated in this first study 
because  only the color structures have been taken into account. The starting 
hypothesis of a maximal singularity
has been found consistent if it appears in a special color structure that cannot
be reproduced by the exchange of one gluon. This singularity is directly
propagated to the other four- and higher $n$-point functions. Although
singularities are induced in all $n$-point functions with $n\ge 4$ the
three-point functions and propagators are protected.  
And, most important, singularities appear whenever the exchanged
momentum between colored subsets of legs of a $n$-point function vanishes. 
In addition, a simple explanation of Casimir scaling is provided.

Already one four-point function even when corresponding to a  very heavy
fundamentally charged field could propogate such a singularity to other
Yang-Mills or matter correlation functions.  Although this is no contradiction to
the decoupling theorem, which holds only in the absence of singularities, it is
nevertheless intriguing. Of course, whether there is then no decoupling, 
or more precisely, how then the decoupling of
infinitely heavy fundamental charges in gauge-invariant quantities takes place
certainly deserves further studies.

Also other related and yet not understood aspects come immediately to one's mind:
How do potential van-der-Waals forces cancel? What about $N$-ality and string
breaking for adjoint charges? Is there a direct relation to dynamical chiral
symmetry breaking?

\section*{Acknowledgements}
We are grateful to the organizers of the {\it Xth Quark Confinement and the Hadron
Spectrum} conference for all their efforts which made this extraordinary event
possible.\\
During the course of this work MM was  funded by the Austrian Science Fund, FWF,
through the Doctoral Program on Hadrons in Vacuum, Nuclei, and Stars (FWF DK
W1203-N16).

\end{document}